# X-Ray Radiation Generation in Heat Engines Using Electron Cooling

Edik A. Ayryan[1], Michal Hnatic[1], Karen G. Petrosyan[2],
Ashot H. Gevorgyan[3], Nikolay Sh. Izmailian[4], Koryun B. Oganesyan[1,4,*]

[1] Joint Institute for Nuclear Research, Dubna, Russia
[2] Institute of Physics, Academia Scinica, Taipei
[3] Yerevan State University, Yerevan, Armenia
[4] Alikhanyan National Science Lab, Yerevan Physics Institute, Yerevan, Armenia

[*] bsk@yerphi.am

**Abstract.** We consider a relativistic quantum heat engine that goes through a thermodynamical cycle consisting of stages involving laser-assisted cooling of electrons and the generation of X-ray radiation. Quantum treatment of the processes makes it possible to obtain the necessary condition and the amount of work extracted from the interaction ingredients, as well as the efficiency of the heat engine. We have also found that the efficiency of the relativistic engine is less than the one for the nonrelativistic case for the same momenta. The obtained results set the limits to the cooling, as well as the intensity of X-ray radiation, in the quantum regime of the interaction of electrons with laser fields.

## 1. Introduction

Interaction of relativistic electrons with magnetic and strong optical fields is of permanent interest to achieve several important goals, such as the cooling of electrons [1, 2] and the generation of X rays [3]. We will approach the problem from the point of view of relativistic quantum thermodynamics [4]. We will build a quantum heat engine [5], employing a particular realization of electromagnetic interactions in the relativistic domain, thus, implementing a relativistic quantum heat engine (RQHE). The stages the RQHE goes through will involve the process of heating the electronic plasma, the free-electron lasing process [6-26], the laser-assisted electron cooling, and the magnetic field controlled motion of the electronic plasma.

Let us consider the following scheme. The electron beam circulates in a ring. It goes through four interaction stages. First, the beam gets accelerated and obtains the energy $E_{k_1}$, whereby the electrons experience a connection with a "thermal bath" with temperature $T_1$. As a result, the beam acquires temperature $T_1$, which is reflected in the transverse motion of the electrons. At the second stage, the electronic beam goes through a process which causes the electrons to adiabatically loose some of their energy. Meanwhile, during this stage, we require that the heat transfer does not take place at all. The electrons work against external forces.
This can be implemented via the interaction of the electronic beam with intense laser fields, making it possible to completely transfer the upper level electron population to a lower level. This is a reversible and controllable process that gives rise to short-wavelength radiation.
After that, the electrons enter the third stage with lower energy , and establish contact with a thermal bath with a colder temperature $T_2$. Effectively, the beam bumps into a strong laser pulse maintained by an optical cavity. As a result, the electrons may emit photons at an optical wavelength. Thus, the beam slows down and returns to its initial low energy level. Hence, we have an isothermal process where the electrons transfer some of their longitudinal energy into optical radiation.

Then, at the fourth stage, the electronic beam adiabatically (without any heat transfer) returns to its initial state. At this stage, work on the system is completed.

## 2. Calculation of the efficiency of RQHE

Let us now turn to the calculation of the efficiency of our RQHE. For the purposes of our discussion here, we present expressions for quantum mechanical internal energy, heat, and work. The expectation value of the measured energy of a quantum system is

$$U = \langle E \rangle = Tr(\rho E) = \sum_i p_i E_i \tag{1}$$

in the energy eigenstates basis, where $E_i$ are the energy levels and $p_i$ are the corresponding occupation probabilities. Then, we can express the internal energy as

$$dU = \sum_i \left( E_i dp_i + p_i dE_i \right), \tag{2}$$

from which we deduce for infinitesimal heat transferred

$$\delta Q \equiv \sum_i E_i dp_i \tag{3}$$

and the work completed

$$\delta W \equiv \sum_i p_i dE_i \tag{4}$$

Note that the above equations imply the first law of thermodynamics $dU = \delta Q + \delta W$.

For calculations of the averages for the quantities presented above, one needs to have an expression for the density matrix. We will assume that all processes taking place in this thermodynamical cycle are in equilibrium.

Having a Hamiltonian for the system and a well-defined temperature, we assume that the density matrix is Gibbsian. There may be several ways to define the density matrix for relativistic cases [31]. We will work out an approach, that is closest to nonrelativistic thermodynamics. As an expression for the density matrix, we postulate, here, the following:

$$\rho \propto \exp(-H/T), \tag{5}$$

where $T$ is the temperature and $H$ is the Dirac Hamiltonian with energy spectrum $E_k = \sqrt{m^2 c^4 + c^2 (\hbar k)^2}$, [32].

Hence, the probability of particles being in a certain energy state is given by

$$p(E_k) \Box\ e^{-E_k/T}. \tag{6}$$

Notice that, in this density matrix, we have a scalar temperature rather than being four-vector temperature, such as that proposed by van Kampen [31] in the momenta distribution. In his remarkable paper, van Kampen pointed out that his distribution can be extended to Bose and Fermi statistics. However, the extension of relativistic expressions to the quantum domain was not made. Hence, we present the above density matrix for our relativistic quantum case and provide the necessary link between relativistic quantum statistical mechanics and relativistic quantum thermodynamics.

Now, we turn to a more detailed description of the stages our RQHE goes through. During *the first stage*, the electronic beam establishes contact with a "thermal bath." At this stage, we have an isothermal process that does not go through a temperature change, that is, $T = T_1$. We assume that, at the end of the fist stage, the beam speeds up in the longitudinal direction and acquires the longitudinal momentum $k_1$. This can be achieved by simply heating up the electronic plasma by guiding it as a whole into a chosen direction. Integrating out the transverse degree of motion and taking into account the quantum nature of the system (although the spectrum is a continuum, it is assumed, that our system can be in either a low-energy, say zero momentum, level state or in an upper level state with a nonzero momentum), we arrive at the following probability of the upper state occupation:

$$p = \frac{1}{1 + e^{E_1/T_1}}, \tag{7}$$

where

$$E_{k_1} = \sqrt{m^2 c^4 + c^2 (\hbar \mathrm{k})^2} - mc^2. \tag{8}$$

There is no work done on or by the system at this stage. Here, we have only heat transferred to the system. During *the second stage*, we assume that there is work completed by the electronic beam without any heat transfer. The system goes through an adiabatic change of its energy from $E_{k1}$ to $E_{k2}$. Let us consider the electronic beam in the quantum regime [33] as a two-level system interacting with two counterpropagating electromagnetic waves having frequencies $\omega_s$ and $\omega_i$ and amplitudes $\varepsilon_{s,i}$. It was shown in [33] that the probabilities $|a_{1,0}|^2$ of being in the upper (lower) level states are

$$\begin{aligned}|a_0(L)|^2 &= |a_0(0)|^2 \sin^2[\chi(L)/2] \\ |a_1(L)|^2 &= |a_1(0)|^2 \cos^2[\chi(L)/2]\end{aligned} \tag{9}$$

where

$$\chi(L) = \frac{\mu}{c\hbar} \int_0^L \varepsilon_s(z) dz,$$
$$\mu = \frac{e^2 \varepsilon_i}{m \omega_i^{1/2} \omega_s^{3/2}}. \tag{10}$$

The equation for $\chi(z)$ is as follows:

$$\frac{\partial^2 \chi}{\partial z^2} = \alpha^2 \sin \chi,$$

$$\alpha = \sqrt{\frac{2\pi n_e e^4 I_i}{\hbar \omega_s^2 \omega_i m^2 c^2}},$$

(11)

Where $L$ is the interaction length in the optical undulator, is the density of electrons, and is the intensity of the laser field.

The solution for the above equations was presented and analyzed in some detail in [33]. Here, we present the values for the length of the optical undulator when one has complete transfer to the lower level:

$$L_m = (2m+1)\alpha^{-1} k K(k),$$

(12)

where m = 0,1,..., $k = \sqrt{1 + |\varepsilon_s(0)|^2 / \hbar \omega_s n_e c}$, and $K(k)$ is the complete elliptic integral. Notice that the regime of complete population transfer is realized, which is analogous to the formation of a π-pulse in a coherent amplifying medium [34].

As a result, during this second stage, the electrons transfer from the energy level to the lower energy level and emit radiation at a short wavelength. The estimates in [33] suggest that an electronic beam with energy $E \approx 2.5$ MeV and density $n_e \approx 3 \times 10^{12} cm^{-3}$ will emit soft X rays with $\hbar \omega_s \approx 100 eV$ in the case of an intraundulator interaction length $L \approx 6$ cm. Let us point out that there is no heat transfer at this second stage. There is only work completed by the system and, as a result, the emission of γ or X rays.

*At the third stage*, there is no work done on or by the system. However, there is a heat transfer that puts the electrons into an even lower energy level. It can be effectively implemented by letting the beam interact with an intense laser field in the manner described in [1,2]. In [2], electronic beam cooling was proposed. The goal was supposed to be achieved in a laser–electron storage ring, where the electrons circulate in a ring and periodically bump into an intense laser pulse being maintained by an optical cavity. A similar scheme was proposed in [1] for a linear collider geometry. We will employ, here, the same idea of the collision of electrons with an intense laser field. We assume that the process is isothermal. Effectively, it can be considered as an interaction with a "thermal bath" with temperature $T_2$. The interaction of the beam with the intense laser pulse causes the electrons to emit photons into all three degrees of freedom. It slows down both the longitudinal and transverse motion of the particles and leads to an effective cooling of the beam. It was estimated in [1] that a 5 GeV electron can lose 90% of its energy in one single passage of a laser pulse with a flash energy of a few *J*. Additional estimates presented in [2] illustrate the results of their laser-assisted electron cooling scheme. Single pass radiated power is given by [2]

$$P = \frac{32\pi}{3} r_e^2 \gamma^2 I,$$

(13)

where $r_e = e^2 / 4\pi mc^2 = 2.82 \times 10^{-15} m$ is the classical radius of the electron and $\gamma$ is the electron energy in units of the resting energy $mc^2$. The energy loss of the electron after passing through the laser pulse is

$$\Delta E = \int P \frac{dz}{2c} = \frac{32\pi}{3} r_e^2 \gamma^2 \frac{E_L}{Z_R \lambda_L},$$

(14)

where $E_L = \int I dx dy dz$ is the laser flash energy, $Z_R$ is the Rayleigh range of the optical resonator, and $\lambda_L$ is the wavelength of the laser. For intense laser pulses, the laser–electron interaction can induce an energy loss which is a sizable fraction of the electron energy $\Delta E \sim E$. Further details can be found in [2].

Finally, *at the fourth stage*, work is complete on the system that should bring the beam back to its initial state. Again, we require that there is no heat transfer at this stage. This can be accomplished by simply applying a magnetic field that will guide the electrons to their starting place. Thus, we have described all four stages of the RQHE. The first stage involves a process of heating the electronic plasma. At the second stage, we have, in essence, a free-electron lasing process. The third stage is simply laser-assisted electron cooling. The last stage is a magnetic field controlled motion of the electronic plasma.

Let us now return to the quantum thermodynamical calculations. The bulk of the work done by the RQHE during the two quantum adiabatic passages at the second and fourth stages is [30]

$$\Delta W = (p_1 - p_2)\left(E_{k_2} - E_{k_1}\right), \tag{15}$$

where $E_{k_1,k_2}$ are the energies of the electronic beam at the beginnings of the second and fourth stages. Here, $p_{1,2}$ are the probabilities of the beam being in their "upper" state with the energies $E_{k_1,k_2}$ as presented above:

$$p_{1,2} = \frac{1}{1 + e^{E_{k_1,k_2}/T_{1,2}}}. \tag{16}$$

Once again, we point out that the beam of electrons is treated as an effective two-level system with a "lower" level corresponding to the zero-momentum particles and an "upper" level for electrons moving with a kinetic energy.

From the above equations, it follows that the work from the engine can be extracted if and only if

$$T_1 > T_2 \left(E_{k_1} / E_{k_2}\right), \tag{17}$$

as follows from the requirement that $p_1 > p_2$. Then, we obtain the efficiency of the RQHE, which is

$$\eta_r = \Delta W / Q_1 = 1 - E_{k_2} / E_{k_1}. \tag{18}$$

Note that our results coincide with those obtained in [30]. This result for the efficiency is general for quantum heat engines considered in [4, 5]. It clearly indicates that the efficiency of the quantum heat engines is less than that for classical Carnot engines:

$$\eta_r < \eta_c \equiv 1 - T_2 / T_1. \tag{19}$$

Actually, it makes sense to compare the RQHE with the quantum heat engine presented in [5]. There is a threelevel maser in their scheme that operates in contact with two thermal baths. There, the system first interacts with the hot thermal bath and enters into the highest level.
Then, contact with a colder bath makes it transfer to a lower state. Eventually, the system emits photons and returns to its initial state. The efficiency obtained is the ratio of the signal frequency to the pumping frequency. In our case, the RQHE first gets "pumped" by the first thermal bath, then does the work by radiating the signal short-wavelength rays, and then cools down while establishing contact with the second thermal bath. The efficiency of the RQHE is the ratio of the signal frequency to the "pumping" frequency (the energy over Planck's constant).

We also notice a difference between the efficiencies for the relativistic and nonrelativistic cases. After a little algebra, it can be derived that

$$\eta_r < \eta_{nr}, \qquad (20)$$

where $\eta_r = 1 - E_2' / E_1'$ and $\eta_{nr} = 1 - E_2^{nr} / E_1^{nr}$ are the efficiencies for the relativistic and nonrelativistic cases, respectively. The corresponding energies are $E_k^r = \sqrt{m^2 c^4 + c^2 (\hbar k)^2} - mc^2$ and $E_k^{nr} = \frac{(\hbar k)^2}{2m}$ .

## 3. Conclusion

In conclusion, we have considered the processes of laser-assisted cooling of electrons and X-ray generation in a quantum heat engine operating in the relativistic domain. The stages the RQHE goes through involve a process of heating the electronic plasma, a free-electron lasing process, laser-assisted electron cooling, and a magnetic field controlled motion of the electronic plasma. Although one needs to apply relativistic quantum thermodynamics for the treatment of the device, the expression for its resulting efficiency is similar to that for nonrelativistic quantum heat engines [5,27-29].
We have also obtained that the efficiency of the relativistic engine is less than that for the nonrelativistic case with the same momenta. Free-electron lasers with or without inversion [35-53] are the main targets of the application of the above described processes. The improvement of conventional schemes of these lasers is still an open question that we will address in a later publication.

**Acknowledgments**

EAA is grateful for the support of RFBR grant 14-01-00628.